# User-Controlled Privacy-Preserving User Profile Data Sharing based on Blockchain

Ajay Kumar Shrestha, Ralph Deters, Julita Vassileva

Department of Computer Science
University of Saskatchewan
Saskatoon, Saskatchewan
aks128@mail.usask.ca, {deters, jiv}@cs.usask.ca

*Abstract*—The tremendous technological advancement in the last few decades has brought many enterprises to collaborate in a better way while making intelligent decisions. The use of Information Technology tools in obtaining data of people's everyday life from various autonomous data sources allowing unrestricted access to user data has emerged as an important practical issue and has given rise to legal implications. Various innovative models for data sharing and management have privacy and centrality issues. To alleviate these limitations, we have incorporated blockchain in user modeling. In this paper, we constructed a decentralized data sharing architecture with MultiChain blockchain in the travel domain, which is also applicable to other similar domains including education, health, and sports. Businesses that operate in the tourism industries such as travel and tour agencies, hotels and resorts, shopping malls etc. are connected to the MultiChain and they share their user profile data via stream in the MultiChain. The paper presents the hotel booking service for an imaginary hotel as one of the enterprise nodes, which collects users' profile data with proper validation and will allow users to decide which of their data to be shared thus ensuring user control over their data and the preservation of privacy. The data from the repository is converted into an open data format while sharing via stream in the blockchain so that other enterprise nodes, after receiving the data, can easily convert them and store into their own repositories. The paper presents an evaluation of the performance of the model by measuring the latency and memory consumption with three test scenarios that mostly affect the user experience. The node responded quickly in all of these cases building a better and more engaging user experience. The paper also proposes a concept of the smart contract in the form of the finite state in the expanding domain of privacy-preserving data sharing and management.

## I. INTRODUCTION

How many of the users are really concerned about the privacy of their own data while using online services? Most people want to have control of their data, what is gathered and how it is used. In reality, this is not the case. For example, in the case of travel, while people are booking their flight or reserving their hotel room, they are also providing their profile data. This is unfortunate that users are transferring the ownership of their own data to those systems and this is one of the problems addressed by this project.

The tourism industries within the hospitality domain usually want to compete successfully and they must do so by using technologies to drive value to all the parties associated with them [1]. Personalization using real-time data about users obtained from different companies is beneficial for their business. This however, can only be achieved by sharing user profile data across trusted companies. This is, however, currently impossible, since informed consent is required to collect, store and use user data by most legislations and there are no existing mechanisms to request and obtain such informed consent from users for secondary use of data (i.e. use that is different from the original purpose for which the data was collected by the company).

User data collected by companies is nearly always kept in centralized servers, which is easy to implement and maintain and is efficient to search and retrieve data. However, these services present an attractive target for hacking and identity theft, and they often get criticized for security issues. There are other risks with centralized third party service provider for the data storage and management of the entire database – the user data may get lost or destroyed or it can be sold to another company whenever the provider gets bankrupt. The new owner of the data may use it for purposes to which the users have not consented and have not been informed. It seems obvious that a secured trustless distributed architecture is needed and a system that guarantees the preservation of user-controlled privacy, as well as it enables users to control if and how their data is shared, with which other companies, for what purpose and under what conditions.

To satisfy these requirements, we have developed a model using blockchain and introduced a sample travel domain in which we created one application owned by a hypothetical company, part of the tourist domain consortium: an online service for hotel booking for an imaginary hotel named Grandee. The different nodes in the blockchain are the participating enterprises in the travel domain, connected with the blockchain on which they can share their user profile data. Blockchain technology can be thought of as a decentralized (peer-to-peer) ledger or database. All user data is encrypted and hashed authenticated in the blockchain system. The content is immutable and verifiable, stored in multitude of peers on the network. The user can decide which data to share and for what reward. This paper reports about an experimental implementation and performance evaluation of the framework for sharing user data among companies owning apps in a travel domain. The performance evaluation is a necessary step to ensure that the platform is scalable and responsive enough for users to interact with in real time and make their decisions about sharing their data.

We have used MultiChain blockchain as a distributed digital ledger in order to provide uneditable private record of all transactions made by the participating users and companies such as travel and tour agencies, hotels and resorts, airlines, shopping malls, etc. Most importantly, being a private blockchain, MultiChain has the potential to replace the traditional centralized databases used in the business model into a decentralized solution, offering more cryptographic auditing features and known identities. Therefore the implementation of the MultiChain blockchain will solve the fraud problems and security issues of the traditional business firms.

Alongside blockchain, smart contracts will aligns the incentives for users to allow their data to be shared and this kind of host profit model is mostly oriented around protecting the uploaded data [2]. There will be an open marketplace where people can provide their resources (profile data) to get more profits just like in the bitcoin mining. The smart contracts for handling user data are a topic of our future work and not discussed in this paper.

The rest of the paper is organized as follows. In Section II, we describe the overview of MultiChain blockchain and brief analysis of the existing architectures with their limitations is covered in Section III. Section IV presents the model that we developed for data sharing in a distributed manner while ensuring users' privacy. In Section V, we evaluate the performance of our model with some experiments. Section VI presents the observation of latency and memory consumption that most affect user experience (UX). In Section VII, we provide the descriptive analysis of the result and in Section VIII, we highlight our future work plan on smart contract to ensure better utilization of user-controlled privacy in the expanding domain of privacy-preserving data sharing and management. Finally, we conclude our work in Section IX.

## II. Background

### A. Blockchain

Where critical assets going through a supply chain, we could use distributed ledger so that we could see where those assets are, what they are doing and we will also have the trust mechanism behind them so that it will be very difficult for the fraudulent agents to inject false goods into the supply chain [2]. Blockchain is a data structure used to create a public or semipublic distributed digital transaction ledger which, instead of resting with a single provider, is shared among a distributed network of computers. Blockchain was first described in the original source code for the digital cash system, Bitcoin [3], but its effect is much broader than just the alternative digital currency. The blockchain is a decentralized database containing every transaction which has ever taken place and is distributed to the edge of the network. Each block aggregates a timestamped batch of transaction to be placed in the chain. There is a cryptographic signature to identify each block. All those blocks refer to the signature of the previous block in the chain, and that chain can be traced back to the very first genesis block created in the chain.

The key idea is that there is no centralized authority to say what is true or what is false, rather multiple independent and distributed nodes come to a consensus on the truth of each new transaction by a process involving proof and voting, after which the transaction is entered into the ledger and thereafter can be accessed by anyone in the future for verification [2]. Computationally, it is impracticable for anyone to go back in order to alter the history because the blockchain has a chronological chain of events, at one particular time one can insert the proof into the public record. The proof is cryptographically protected so only those who got the key can see it and so there is no chance of fraudulence [2].

The Blockchain is on its way to really transform our society from trust-based to truth-based trust-less society. The idea behind the blockchain can be used to store data in different areas. Many financial industries including banks are now working on incorporating blockchain technology as distributed ledgers for their transactions. The internet has now been flooded with the ideas playing around the blockchain, emphasizing it to be the next big thing.

There are three categories of the blockchain. One is public in which anyone can participate in the chain and contribute to the consensus-making process. Another one is the consortium in which pre-selected nodes control the consensus process. And the final type is the private blockchain, which has a closed community storing the transactions that are of interest to only those private participants present in the chain. In our work, we have used a private blockchain in the form of MultiChain which is explained in the next section. Most importantly, MultiChain supports streams that can be exploited to send and receive larger amount of data with a combination of symmetric and asymmetric cryptography.

We have connected different tourism industries including hotel booking system to the MultiChain and they share their user profile data via stream in the blockchain. The permission to send, receive and publish the stream for different nodes can be granted partly or wholly, or restricted fully in that private blockchain.

### B. MultiChain

MultiChain is an off-the-shelf private blockchain that provides the privacy and control required in an easy to configure and deploy package [4]. It supports UNIX and Windows servers and comes up with a rich JSON-RPC API for easy integration with existing systems. Unlike any other blockchains, MultiChain solves the problems of mining, privacy, and openness via integrated management of user permissions, thereby three folding the core aims [4]:

- To ensure that the blockchain's activity is only visible to chosen participants,

- To introduce controls over which transactions are permitted, and

- To enable mining to take place securely without proof of work and its associated costs.

Once a blockchain is private, problems relating to scale are easily resolved, since the chain's participants can control the maximum block size. In addition, as a closed system, the blockchain will only contain transactions which are of interest to those participants. Basically, MultiChain allows the user to set all of blockchain's parameters in a configuration file.

### C. Smart Contracts

Blockchain coupled with smart contract technology removes the reliance on the central system between the transaction parties. Basically, smart contracts are stored on the Blockchain, which all the connected parties in the network have a copy of. The Smart contract is an important piece of software that

- Stores the rules which negotiate the terms of the contract,

- Automatically verifies the contract, and

- Executes the agreed terms.

The smart contract can execute agreed stored process when triggered by an authorized or agreed event just like traditional systems. All contract transactions are stored in chronological order for future access along with the complete audit trail of events. If any party tries to change a contract or transaction on the Blockchain, all other parties can detect and prevent it. If any party fails, the system continues to functions with no loss of data or integrity. It therefore creates a single large secure computer system logically, without the risks, costs and trust issues of a centralized model. Thus the smart contracts represent the rules that manage the process of storing and accessing data in the Blockchain regarding rights of access, type of data, constraints etc. They can be used to handling the storing and accessing of user profile data.

The next section introduces existing systems for sharing user data and their limitations.

## III. RELATED WORKS

Existing user data sharing models at an enterprise level are networked information systems allowing creating and storing user profile data, and making it accessible for others with special agreement. Centralized architectures (user model servers) are predominant because client-server architectures are well established technology, efficient and scalable. In fact, the physical storage of user data is not essential. There are cases in which the user data is stored in distributed storage spaces, but the schema is kept centrally [13], these are still centralized user models.

There are examples of user data sharing systems are experimental frameworks developed in Academia, aiming to achieve interoperability of distributed user models. Different architectures exist. For example, Mypes [5] has a centralized server such as where the representation schemas of different user models/profiles is "translated" to a standard one using ontology mapping. There have been also data management systems proposed, collaborative repositories such as Wikidata [6], etc. Almost all of these systems implement different specific architectures and their evaluation is based upon different non-functional requirements, such as efficiency, scalability, or reliability [7]. However, the technical performance of a data sharing system alone does not guarantee the practicality of the systems. The centralized architecture, at most cases, doesn't collect and share the diverse fragments of user data coming from the enormous autonomous and independent entities (applications, agents, devices, sensors, services) comprising the service-oriented, mobile and ubiquitous computing environment [14]. Often the centralized user modeling technique has a predefined point of access that leads to the central point of failure. Replication of the data via mirroring the servers could be a solution, but that usually comes with high communication cost as well. Therefore, decentralized approaches for user modeling hold more promise to overcome the limitations brought by the centralized user modeling architecture.

In [8], the authors present a distributed architecture for sharing and re-using multi-application life logs. All the life logs from different systems are gathered by agents, which then forward the information to a central broker which is responsible for user modeling that comprises of request analysis, source selection, source connection, semantic mapping, data integration and response transformation.

Even in the IOT domain, there are systems such as MobiTribe [9], which has distributed model. However, it uses a centralized content management system as a moderator for the exchange of information between the devices and the applications. PersonisAD [10] is another active, distributed, scrutable model that gathers information from different sensors associated with different users and combines their preferences in order to provide a richer experience.

Like in [11], the distributed user model is represented by single method standalone agents which store a single attribute of a user model within the holistic vector (stored in a server). In [15], [16], the model is decentralized, held by different agents, and information is gathered from different agents only temporary, for a given purpose of adaptation.

So we see that all existing approaches including those that are distributed involve some form of centralization: it is either permanent, where data from different sources is brought together and aligned semantically, or is temporary, depending on the request and purpose. Furthermore, it is observed that accommodating the conflicting interests among the users is not separable from the architectural design that applies optimization of the specific system properties and involves trade-offs with the participants' autonomy [7]. Even in some (structured) peer-to-peer architectures such as Chord [12], participants are not given the privilege to connect to the other peers of their choice, but rather they have to store data with arbitrary other peers.

The next section describes the working of our model which ensures decentralized sharing of user profile data among an enterprise consortium for testing the usefulness of our model, for the travel domain which includes the hospitality domain.

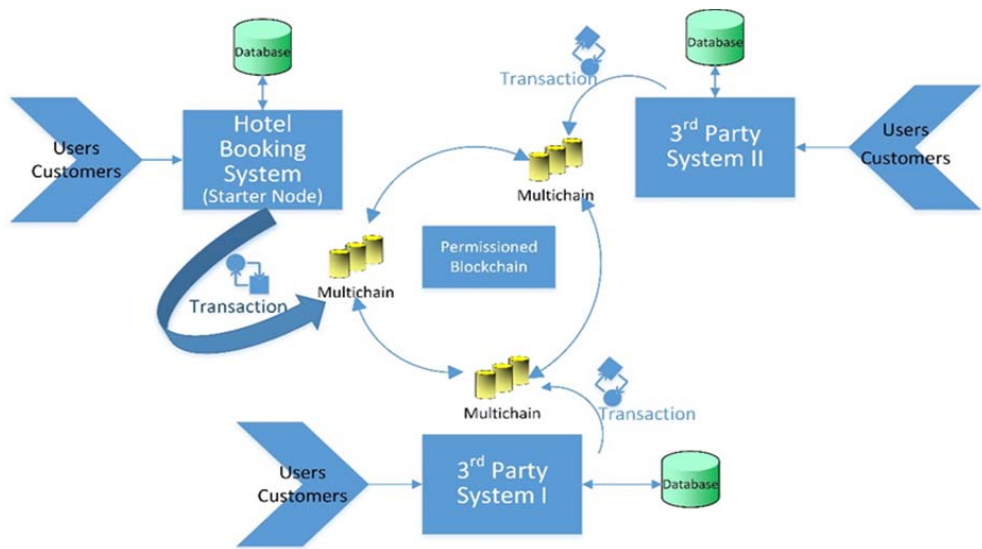

Fig. 1.   Basic user-controlled privacy preserving architecture based on MultiChain for data sharing.

## IV.   METHODOLOGIES AND DISCUSSION

Fig. 1 explains our model. We developed and deployed a general hotel booking system in one UNIX machine. The machine has MultiChain in it and acts as one of the nodes that collect clients' data with proper validation and sanitization. Later on, we used other UNIX and Windows machine for the performance evaluation, which is explained in Section VI. The web application is developed in PHP with Apache server and MySQL as backend. Users create their profile in the hotel booking system on the first node (Node1) in the blockchain by registering their information and choosing which of their data can be shared with third parties. The user profile data includes name, nationality, contact number, purpose of visit and the dates of stay. The data stored in the repository is converted into an open data JSON format, which can be published in the MultiChain via stream. The stream in MultiChain is used for general data storage and retrieval. Other nodes owned by other companies, e.g. Saskatoon Travel and Tours, and Saskatoon Shopping Mall also have Multichain in their system. They get the address and are given permission to be in the closed network of the blockchain. Public key encryption is an underlying technology of MultiChain, so all the connected nodes generate their own pair of public addresses and private keys.

The MultiChain restricts blockchain access to a list of permitted users, by expanding the "handshaking" process that occurs when two blockchain nodes connect [4]:

As shown in Fig. 2, we first created a Multichain in the Hotel Booking system for the first node (Node1) in the blockchain. By default, this node acted as an Admin which could further grant other nodes to be admins too. The permissions for other nodes (connect, send, receive, issue, create, mine, admin, activate) will be set by this node in our case, but it can be made  true for all nodes while setting chain parameters as shown in Fig. 3, which gives the setting for the basic and global blockchain parameters.

The multichain daemon was created exploiting the following command with the chain name model:

multichain-util create model

multichaind model –daemon

Fig. 2.   All connected nodes as seen from Node1-Grandee Hotel.

This created the MultiChain Core Daemon, build 1.0 alpha 27 protocol 10007 such that other nodes can connect to this node using command: multichaind model@[ip]:[port]

(e.g. multichaind model@192.168.204.132:8353).

We then created other two nodes: Node2 and Node3 representing Saskatoon Travel and Tours, and Saskatoon Shopping Mall, respectively as independent imaginary companies in the travel domain. The creation of the nodes offered the individual addresses for those new nodes which were acknowledged by the first node in order to grant a "connection" permission to them into the MultiChain since it is the private blockchain. Back on the first server, we added connection permissions for other node addresses as:

multichain-cli model grant [address] connect, send, …

This is the first step in creating the blockchain. While granting the connection permission, further other permissions can also be set for other nodes. As shown in Fig. 4, Node2 (Saskatoon Travel and Tour) is given connect, send, receive, issue, create, mine, activate and admin permissions, and Node3 (Saskatoon Shopping Mall) is given all except admin and activate. This means Node2 in the blockchain could be able to act as admin but Node3 couldn't. We further created other 7 nodes to evaluate the system performance (presented in Section V). Fig. 5 is the snapshot of how the enterprise created the stream containing user profile data and Fig. 6 shows how to publish the stream with uploading the items containing user profile data into the stream for sharing them to other consortium enterprises nodes.

```
#Basic chain parameters
chain-protocol = multichain    # Chain protocol
chain-description = MultiChain model    # Chain Desc
root-stream-name = root    # Root stream name
root-stream-open = true    # Allow anyone to publish in root stream
chain-is-testnet = false    # Content of the 'testnet' field of API
responses, for compatibility.
target-block-time = 15    # Target time between blocks (transaction
confirmation delay), seconds. (5 - 86400)
maximum-block-size = 8388608    # Maximum block size in bytes.
(1000 - 1000000000)

#Global permissions
anyone-can-connect = false    # private blockchain.
anyone-can-send = false    # transaction signing is not restricted by
address.
anyone-can-receive = false    # transaction outputs are restricted by
address.
anyone-can-receive-empty = true    #without permission grants, asset
transfers and zero na$
anyone-can-create = false    # selected can create new streams.
anyone-can-issue = false    # selected can issue new native assets.
anyone-can-mine = false    # selected can mine blocks (confirm
transactions).
anyone-can-activate = false    # selected can grant or revoke connect, send
and receive permissions.
anyone-can-admin = false    # selected can grant or revoke all
permissions.
support-miner-precheck = true    # Require special metadata output with
cached scriptPubKey for input, to support advanced mine$
```

Fig. 3.    Setting for the basic and global chain parameters.

Current Permissions

| Label | Grandee Hotel |
|---|---|
| Address | 1GNcxezAZfivdopnxDf6X4RDDXN4jzG4b6SDwr (local) |
| Permissions | mine, admin, activate, connect, send, receive, issue, create |
| Label | Saskatoon Travel and Tours |
| Address | 1QzCap7vKVoyvRuoHa6wSgUr46x7KPN3u1U3FK |
| Permissions | mine, admin, activate, connect, send, receive, issue, create |
| Label | Saskatoon Shopping Mall |
| Address | 1EgR7kbiAzAN1HmzVrod8saPGNg5oD9yg6hmEa |
| Permissions | mine, connect, send, receive, issue, create |

Fig. 4.    Permissions granted to connected nodes as seen from Hotel Booking System (Node1).

Fig. 5.    Creation of the stream.

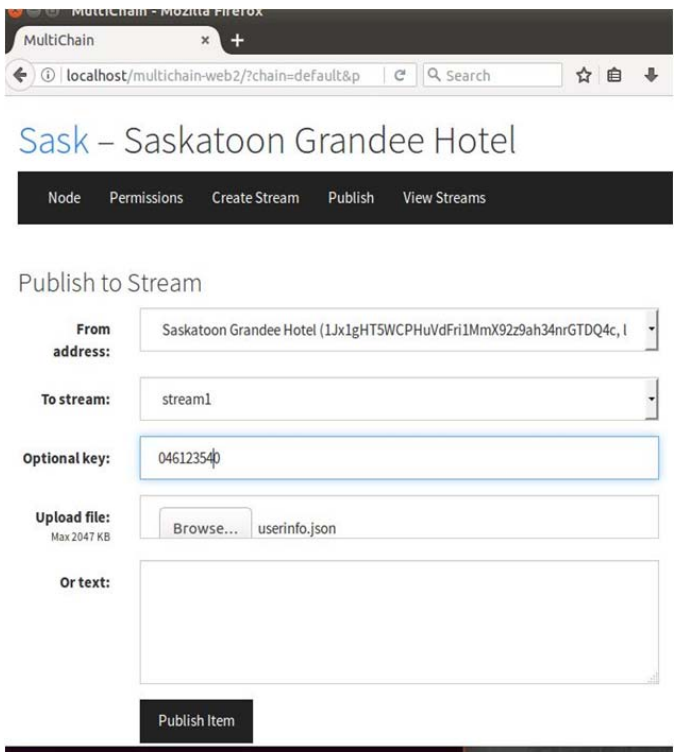

Fig. 6.    Publishing stream of items.

## Subscribed streams

| Name | root |
|---|---|
| Created by | Grandee Hotel (1GNcxezAZfivdopnxDf6X4RDDXN 4jzG4b6SDwr) |
| Items | 6 |
| Publishers | 6 |

| Name | User data-1 |
|---|---|
| Created by | Grandee Hotel (1GNcxezAZfivdopnxDf6X4RDDXN 4jzG4b6SDwr) |
| Items | 2 |
| Publishers | 1 |

| Name | User data-2 |
|---|---|
| Created by | Grandee Hotel (1GNcxezAZfivdopnxDf6X4RDDXN 4jzG4b6SDwr) |
| Items | 1 |
| Publishers | 1 |

Fig. 7.    List of streams created by Node 1.

### Stream: User data-1 – 4 of 4 items

| Publishers | Grandee Hotel (1GNcxezAZfivdopnxDf6X4RDDXN4jzG4b6SDwr) |
|---|---|
| Key | |
| Data | userinfo.json (2 KB) |
| Added | 2017-05-28 01:08:38 GMT |

| Publishers | Grandee Hotel (1GNcxezAZfivdopnxDf6X4RDDXN4jzG4b6SDwr) |
|---|---|
| Key | 012012 |
| Data | chain1_info (1 KB) |
| Added | 2017-05-28 01:08:07 GMT (confirmed) |

| Publishers | Grandee Hotel (1GNcxezAZfivdopnxDf6X4RDDXN4jzG4b6SDwr) |
|---|---|
| Key | |
| Data | testdata.json (2 KB) |
| Added | 2017-05-25 08:14:17 GMT (confirmed) |

| Publishers | Grandee Hotel (1GNcxezAZfivdopnxDf6X4RDDXN4jzG4b6SDwr) |
|---|---|
| Key | |
| Data | empdata.json (1 KB) |
| Added | 2017-05-25 05:14:40 GMT (confirmed) |

Fig. 8.    Different files in the stream.

Fig. 7 shows the list of the streams created by the Node 1-Grandee Hotel. The first node was the Hotel Booking System which basically collected the users' data while reserving rooms in the hotel. This information is useful to other tourism enterprises like the shopping mall, and travel and tours so that they can provide attractive offers to users during their stay. So the collected useful information of user profile is to be shared among the enterprise consortium as per the predefined agreed terms.

The collected data at Node1 were added as items into the stream. Tthe added files into the stream were then published and distributed into all the nodes. All files present in the stream can be seen in the form of items as in Fig. 8. Only the nodes with the receive permission can view the contents from the streams.

All other nodes in the network can easily convert and store the received stream file into their own repositories. In fact, every node in the MultiChain blockchain can have access to any stored raw data. In order to resolve this issue of confidentiality, data is encrypted before being put into the chain. The three blockchain streams with a combination of symmetric and asymmetric cryptography have been used [4]:

*a) Pubkeys stream:* It is used by participants to distribute their public keys under the RSA public-key cryptography.

*b) Items stream:* It is used to publish large pieces of data, each of which is encrypted using symmetric AES cryptography scheme.

*c) Access stream:* It provides data access. For each participant who should see a piece of data, a stream entry is created which contains that data's secret password, encrypted using that participant's public key.

We have combined Multichain and off-blockchain repository to create a data sharing and management model focused on security and privacy. The next section will give the insights of evaluating the performance of our model.

## V. PERFORMANCE EVALUATION

It is very important to evaluate the system performance by analyzing the performance metrics that mostly affect user experience (UX). We evaluated the performance of the system by carrying out successive experiments on the freshly created nodes. We set three goals – to find out:

*1)* How long it takes the enterprise in the form of multichain node to get connected to the network?

*2)* How long it takes the enterprise in the form of multichain node to respond to actions (like starting stream, viewing a stream item, loading or publishing the items into the stream)?

*3)* How much memory the node consumes when blockchain Daemon gets started.

### A. Expected Outcomes

We expected that the node should respond quickly in all of the cases mentioned before for our model, building a better and more engaging UX. We tracked the network latency values in three different scenarios involving different numbers of nodes, and expected them to be within the acceptable range (below 500 milliseconds). The consumption of memory is the third performance metric that we expected to be as low as an acceptable value around 50 MB, so that the system can operate and be handled in an efficient way.

### B. Workload Justification

The theoretical peak bandwidth of a network connection is fixed as per the technology used. However, the actual number of packets to be sent over network is affected by higher and lower latencies. Excessive latency prevents data from filling the network pipe, thus decreasing throughput and limiting the maximum effective bandwidth of a connection. Therefore we set our goal of the evaluation to retrieve the latency in each case which is explained in the next Experimental Setup section.

### C. Experimental Setup

To evaluate the implementation prototype, we performed an evaluation plan to simulate the real-world interactions. The evaluation involved three scenarios to simulate different levels of concurrency while monitoring latencies. The three scenarios are shown in Table 1.

We carried out the experiments in the Windows and the UNIX machines. We stopped all extra processes except the basic OS processes to run in the background alongside the Multichain daemon so as to ensure that no other process would affect our experiments. The detailed list of the used computer system for fresh nodes is given in Table 2.

TABLE I.  TEST SCENARIO DESCRIPTION

| Scenario | Descriptions |
|---|---|
| S1 | Two enterprise nodes connected |
| S2 | Three enterprise nodes connected |
| S3 | Eight enterprise nodes connected |

TABLE II.  MACHINES DESCRIPTION

| Nodes | System Description |
|---|---|
| N1 | Windows 10 Home; Lenovo ThinkCneter M900 Signature Edition Intel® Core™ i7-6700 CPU @ 3.4 GHz 3.41GHz Memory 32 GB 64-bit OS, x64-based processor |
| N2 | Window 10 Pro; HP Intel® Core™ i5-3427U CPU @ 2.3 GHz Memory 4 GB 64-bit OS, x64-based processor |
| N3 | Windows 7 Professional Education Sony Vaio Interl®Core™i3-2310M CPU @ 2.1 GHz Memory 4GB 64-bit OS, x64-based processor |
| N4 | Windows 10 Education; Lenovo Intel® Core™ i7-3770 CPU @ 3.4 GHz 3.40GHz Memory 32 GB 64-bit OS, x64-based processor |
| N5 | Window 8 Enterprise, Lenovo ThinkCneter M900 Signature Edition Intel® Core™ i7-4770 CPU @ 3.40 GHz 3.40GHz Memory 32 GB 64-bit OS, x64-based processor |
| N6 | Window 10 Home, Lenovo ThinkCneter M900 Signature Edition Intel® Core™ i7-6700 CPU @ 3.4 GHz 3.41GHz Memory 16 GB 64-bit OS, x64-based processor |
| N7 | Window 8 Professional Mac Intel® Core™ i7-2640 CPU @ 2.8 GHz 2.8GHz Memory 4 GB 64-bit OS |
| N8 | Window 10 Home, Lenovo ThinkCneter M900 Signature Edition Intel® Core™ i7-6700 CPU @ 3.4 GHz 3.41GHz Memory 16 GB 64-bit OS, x64-based processor |

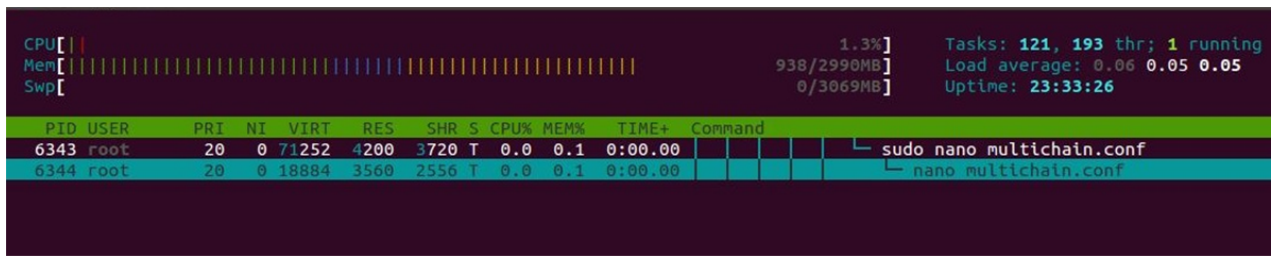

Fig. 9.   Memory status at normal instance.

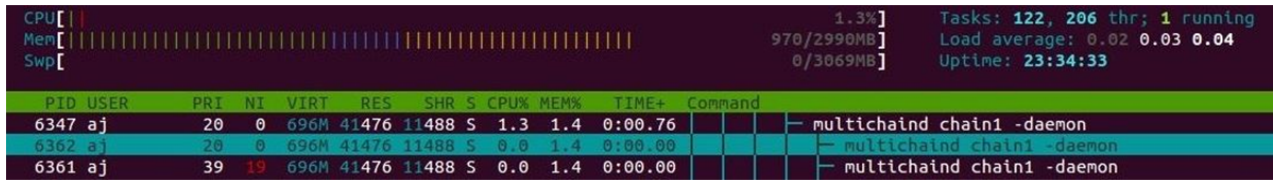

Fig. 10.  Memory status after multichain daemon started.

.

*1) Node Setting:*

The experiment was carried out on the newly created Multichain nodes. Since the MultiChain uses the cryptography mechanism, it restricts block index and chainstate access to the list of permitted users; so we created blockchain nodes as fresh ones. The block index maintains information for every block, and where it is stored on disk. The chain state maintains information about the resulting state of validation as a result of the currently best-known chain. Basically, node parameters have been set up as stated before in order to store the key-value pairs of all the block and state hashes. More specifically, the nodes use the following interaction protocol:

*a)* Each node presents its identity as a public address on the permitted list.

*b)* Each node verifies that the other's address is on its own version of the permitted list.

*c)* Each node sends a challenge message to the other party.

*d)* Each node sends back a signature of the challenge message, proving their ownership of the private key corresponding to the public address they presented.

If either node is not satisfied with the results, it aborts the peer-to-peer connection.

## VI.   Observations

Our experiment focuses on observing the values of two factors: latency and memory consumption. To observe the effect of the multichain core daemon being stopped and reconnected into the network, we made the scripts that run with the gap of 1 minute for every new observation using the following multichain commands:

multichain-cli model stop

multichaind model daemon

We observed the latency from the first node Node1, when it got connected to another single node N2 for scenario S1, other two nodes N2 and N3 for scenario S3 and other seven nodes N2-N7 for scenario S3 in a total of 20 observations.

For S3, first, we recorded the latency from Node1 to connect it with the other 7 nodes in the network and then finally took their average to get the mean latencies for connecting 7 different nodes from Node1.

Furthermore, we also carried out another experiment to observe the memory consumption for the nodes when the corresponding multichain core daemon got started on that particular node. The total of five observations was carried out, one of which is highlighted as in Fig. 9 and Fig. 10, showing the total memory usage during the pre and post activation of multichain Daemon. Next section will give the detailed analysis of the results obtained during the observations.

## VII.   Result Analysis

The data on latency for the first part of the observation is shown in Table 3 and Fig. 11. All the scenarios have the minimum and maximum latency around 100ms and 150ms respectively, giving the average latency time around 125ms, which is within the acceptable margin. It can be concluded that there is no scalability limit in terms of node count, because each node doesn't need to connect to every other to create a fully connected peer-to-peer network.

Moreover, for all the node catch-up time, it can be concluded that new nodes joining the chain have to replay all transactions from the beginning, and so it can take them significant time before they are up-to-date. The exact amount of time will depend on how many blocks and transactions are in the chain. Our experiment had been carried out with only 10 streams with 100 items in total which was below 100MB. It is because we were only concerned with the latency. In addition to that, since no smart contract is running in the Multichain blockchain unlike other blockchains such as Ethereum, there is no execution of any automated program for every message on every blockchain node. That surely contributed to the low latency that we observed here. Also, today's main issue within distributed applications is not the TX cost as people can handle a cent, but probably is the latency as people want 200ms, not 2s and multichain nodes are really fast in making a connection to the existing blockchain.

TABLE III.    LATENCY (MS) SUMMARY FOR THREE SCENARIOS

| Scenarios | N | Min | Max | Avg. | SD |
|---|---|---|---|---|---|
| S1 | 20 | 85 | 159.5 | 122.57 | 19.32 |
| S2 | 20 | 80 | 156 | 126.2 | 24.24 |
| S3 | 20 | 106.86 | 144.7 | 127.22 | 11.01 |

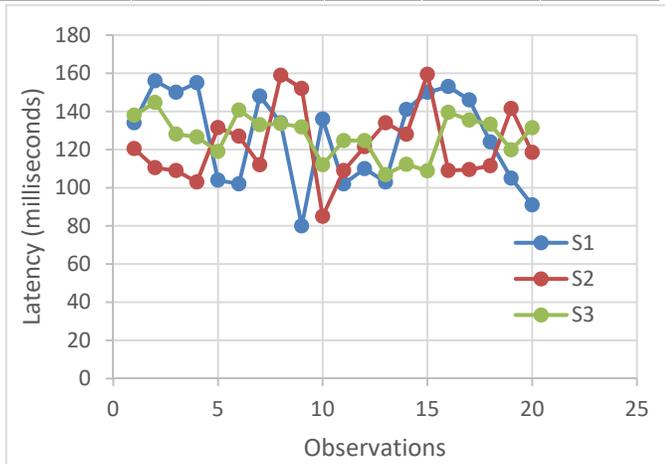

Fig. 11. Latency test result in chart for three scenarios.

TABLE IV.    TESTING MEMORY CONSUMPTION

| Memory usage (MB) | | |
|---|---|---|
| Initial | Later-<br>daemon started | Total<br>multichain -daemon |
| 938 | 970 | 28 |
| 938 | 970 | 28 |
| 938 | 970 | 28 |
| 938 | 970 | 28 |
| 938 | 970 | 28 |

Furthermore, we also analyzed the memory consumption for nodes when their multichain core daemons started. We carried out 5 successive observations as provided in Table 4 and found that latency stayed the same - 28MB. So it can be concluded that the memory usage is not huge to operate our model with the Multichain blockchain. Moreover, it is also based on the number of the unspent transaction. In fact, there is also around 300 bytes held in memory for each block in the chain. Therefore, if the node is subscribed to millions of streams, then that would definitely increase memory usage. However, our model has focused on storing the user profile data and even 1 million of those data will have a size just around 100 MB. So this model is very effective in terms of quick start, quick response and fewer memory consumptions.

## VIII.    FUTURE WORK

The major concern with online applications including our model is that users have their own characteristics, e.g. preferences, interests etc. which comprise the user model of the system. Users are often required to grant a set of permissions upon sign up. In our model, to ensure the user-controlled privacy, the permissions are granted indefinitely and the only way to alter the agreement is by opting-out. In future, we will extend our model to include smart contracts so that access-control policies would be stored securely on the blockchain while retaining the same user-interface and only the user is allowed to change the permissions. Through smart contract which will be in the form of a piece of software, we will automate the functionality that supports the user-controlled privacy: whom to share the users data with, how long the data is to be kept, and how will the user be incentivized for providing access to their data, how to ensure compliance with the contract once the data has been sold to other third parties etc. Since the MultiChain blockchain doesn't have smart contract feature, alternatively we can have a finite state machine coupled with the blockchain, so at any given state, the user can alter the set of permissions (state1) and withdraw access (state2) to previously collected users' profile data.

## IX.    CONCLUSION

In summary, we have performed an experimental study of the use of blockchain in the user modeling and evaluated the system performance by observing the latency and memory consumption. To share the users' profile data in a decentralized fashion, the concept of streams from the Multichain has been successfully interpreted by taking as an example the case of travel domain. This eliminates the single point of failure and centrality issues which are often present in the centralized user model servers. This blockchain based user model is not just limited to travel domain but also applicable to other similar domains such as education, health, sports etc. The paper has evaluated the system performance of our implementation and it met our expectations in terms of the latency and memory consumption. In our future work, we will use the concept of the smart contract to allow the user decide for how long his/her profile data is going to be stored in the system and, with whom his/her data is going to be shared for, and how will he/she get rewarded for sharing. The future model with smart contract will ensure that active ownership and control of user data stays with the users.

REFERENCES

[1] Cassidy C., Chae B. Consumer information use and misuse in electronic business: An alternative to privacy regulation // Information Systems Management. – 2006. – No. 23. – P. 75– 87.

[2] Shrestha A. K. and Vassileva J. (2016). Towards decentralized data storage in general cloud platform for meta-products. In the Proceedings of the International Conference on Big Data and Advanced Wireless Technologies (BDAW '16). ACM, New York, NY, November 10 - 11, 2016.

[3] GitHub, original-bitcoin/main.h at master trottier/original-bitcoin https://github.com/trottier/original-bitcoin/blob/master/src/main.h#L795-L803. (Accessed Feb 10, 2017).

[4] Greenspan, G. MultiChain Private Blockchain — White Paper; 1st ed.; 2015.

[5] Abel, F., Herder, E., Houben, G.J., Henze, N., Krause, D.: Cross-system user modeling and personalization on the social web. User Modeling and User-Adapted Interaction 23(2-3), 169–209 (Apr 2013).

[6] Vrandecic, D. and Krotzsch, M. "Wikidata: A free collaborative knowledge- base," Commun. ACM, vol. 57, pp. 78–85, Sept. 2014.

[7] Davoust, A. 2015. Decentralized Social Data Sharing. Carleton University.

[8] Iyilade, J., Vassileva, J. 2013. A decentralized architecture for sharing and reusing lifelogs. In UMAP Workshops..

[9] Thilakarathna, K., Petander, H., Mestre, J., Seneviratne, A.: {MobiTribe}: Cost Efficient Distributed User Generated Content Sharing on Smartphones. IEEE Transactions on Mobile Computing PP(99), 1–1 (2013).

[10] Assad, M., Carmichael, D. J., Kay, J., Kummerfeld, B. : PersonisAD: Distributed, Active, Scrutable Model Framework for Context-Aware Services. In: LaMarca, A., Langheinrich, M., Truong, K.N. (eds.)

Pervasive Computing, pp. 55–72. No. 4480 in Lecture Notes in Computer Science, Springer Berlin Heidelberg (Jan 2007).

[11] Dim, E., Kuflik, T. 2012. User models sharing and reusability: a component-based approach In: UMAP Workshops.

[12] I. Stoica, R.Morris, D. Liben-Nowell, D. R. Karger, M. F. Kaashoek, F. Dabek, and H. Balakrishnan, "Chord: a scalable peer-to-peer lookup protocol for in- ternet applications," Networking, IEEE/ACM Transactions on, vol. 11, no. 1, pp. 17–32, 2003.

[13] Carmagnola, F., Cena, F., and Gena, C. (2011). "User Model Interoperability: a Survey". User Model User - Adapted Interaction, pp. 1-47, Springer, Netherland.

[14] Dolog, P. and Vassileva, J. (2005). Decentralized, Agent-based and social approaches to User Modeling. In: Workshop DASUM-05, at the 9th International Conference on User Modeling (UM '05). Edinburgh, Scotland.

[15] Niu X., McCalla G. I., and Vassileva J. (2004). Purpose-based Expert Finding in a Portfolio Management System. Computational Intelligence Journal, 20 (4), 548-561.

[16] Vassileva J., McCalla G., Greer J. (2003). Multi-Agent Multi-User Modeling, User Modeling and User-Adapted Interaction, 13:(1), 179-210.